\documentclass[reprint,amsmath,amssymb,aip,apl]{revtex4-2}

\usepackage{graphicx}
\usepackage{amsmath}
\usepackage{bm}
\usepackage[T1]{fontenc}
\usepackage[utf8]{inputenc}

\begin{document}

\title{High-stability offset-frequency locking of two lasers using a balanced filter discriminator}

\author{Sang Eon Park}
\email{parkse@kriss.re.kr}
\author{Meung Ho Seo}
\author{Young-Ho Park}
\author{Hyun-Gue Hong}
\author{Sang-Bum Lee}
\author{Sangwon Seo}
\author{Jae Hoon Lee}
\author{Seji Kang}
\author{Taeg Yong Kwon}
\affiliation{Korea Research Institute of Standards and Science (KRISS), Daejeon 34113, Republic of Korea}

\date{\today}

\begin{abstract}
We demonstrate a high-stability laser offset-frequency locking technique based on a balanced filter discriminator. The beat note between two 852~nm external-cavity diode lasers is down-converted in two parallel arms using local-oscillator frequencies placed symmetrically around the desired offset frequency. After low-pass filtering and RMS detection, differential subtraction of the two detector outputs produces a dispersive frequency-error signal with a zero crossing primarily defined by the reference local-oscillator frequencies. This balanced configuration reduces sensitivity to common beat-power fluctuations and can improve the effective error-signal signal-to-noise ratio. The system was implemented for an 8.653~GHz offset corresponding to the cesium repumping frequency difference used in our laser-cooling setup. Measurements with different low-pass filters reveal a trade-off between discrimination sensitivity and feedback bandwidth. With an SLP-1.9+ filter, the locked beat frequency reached a fractional instability of $4\times10^{-15}$ at 10~s when referred to the 852~nm optical carrier. The residual dependence on photodetector optical power was also characterized, showing that amplitude-to-frequency conversion remains small in the optimized differential configuration. This approach provides a practical frequency-only offset-locking method for atomic-physics experiments requiring stable and tunable microwave-scale laser frequency offsets.
\end{abstract}

\maketitle

The precise control of laser frequencies and their mutual frequency differences is essential in atomic, molecular, and optical physics and precision frequency metrology. Laser-cooling experiments,\cite{Phillips1998} atomic clocks,\cite{Wynands2005,Lee2021APL,Lee2017IEEE,Ludlow2015} atom interferometers,\cite{Peters1999,Kasevich1991} and frequency-comb--referenced laser systems\cite{Jones2000,Udem2002} often require one laser to be stabilized at a well-defined offset from another reference laser. In alkali-atom experiments, for example, cooling, repumping, optical-pumping, and Raman lasers must maintain stable frequency separations over timescales ranging from short experimental cycles to long unattended operation. In optical-frequency-comb--referenced systems, this concept can be extended to stabilize auxiliary lasers to selected comb modes, allowing large optical-frequency separations to be bridged through appropriate down-conversion or frequency-division stages.\cite{Jones2000,Udem2002}

Offset-frequency locking is commonly realized with optical phase-locked loops,\cite{Yim2014,Yim2019,Xu2012} electronic delay-line discriminators,\cite{Schunemann1999} filter-based frequency discriminators,\cite{Ritt2004,Schilt2008,Li2022} digital counting schemes,\cite{Hughes2008} or other standalone electronic approaches.\cite{Shalaby2026} Optical phase-locked loops provide tight relative phase control and are indispensable for applications such as Raman atom interferometry, but they require high servo bandwidth and careful loop optimization.\cite{Kasevich1991,Yim2014,Yim2019,Xu2012} Frequency-only offset-locking methods are often robust when phase coherence is not required. For high-stability operation, however, the achievable frequency stability can be limited by the discrimination slope, residual amplitude-to-frequency conversion, and the long-term stability of the discriminator response.\cite{Ritt2004,Li2022} Some implementations therefore employ normalization channels or active power-stabilization circuits to stabilize the discriminator response.\cite{Ritt2004,Li2022}

Here, we demonstrate a balanced filter discriminator for high-stability laser offset-frequency locking. The beat note between a master laser and a slave laser is divided into two down-conversion arms driven by local-oscillator frequencies placed symmetrically around the desired offset frequency. After parallel low-pass filtering and RMS detection, the two detector outputs are subtracted to produce a dispersive error signal. The zero crossing is primarily determined by the local-oscillator frequencies, while the differential configuration reduces common-mode beat-power fluctuations and can improve the effective error-signal signal-to-noise ratio. This architecture provides a practical frequency-only offset-locking method without requiring a separate normalization signal or active beat-power stabilization.

We validate the method using an 8.653~GHz offset lock between two 852~nm external-cavity diode lasers and characterize its stability for a cesium laser-cooling application.

\begin{figure}[t]
\centering
\includegraphics[width=\columnwidth]{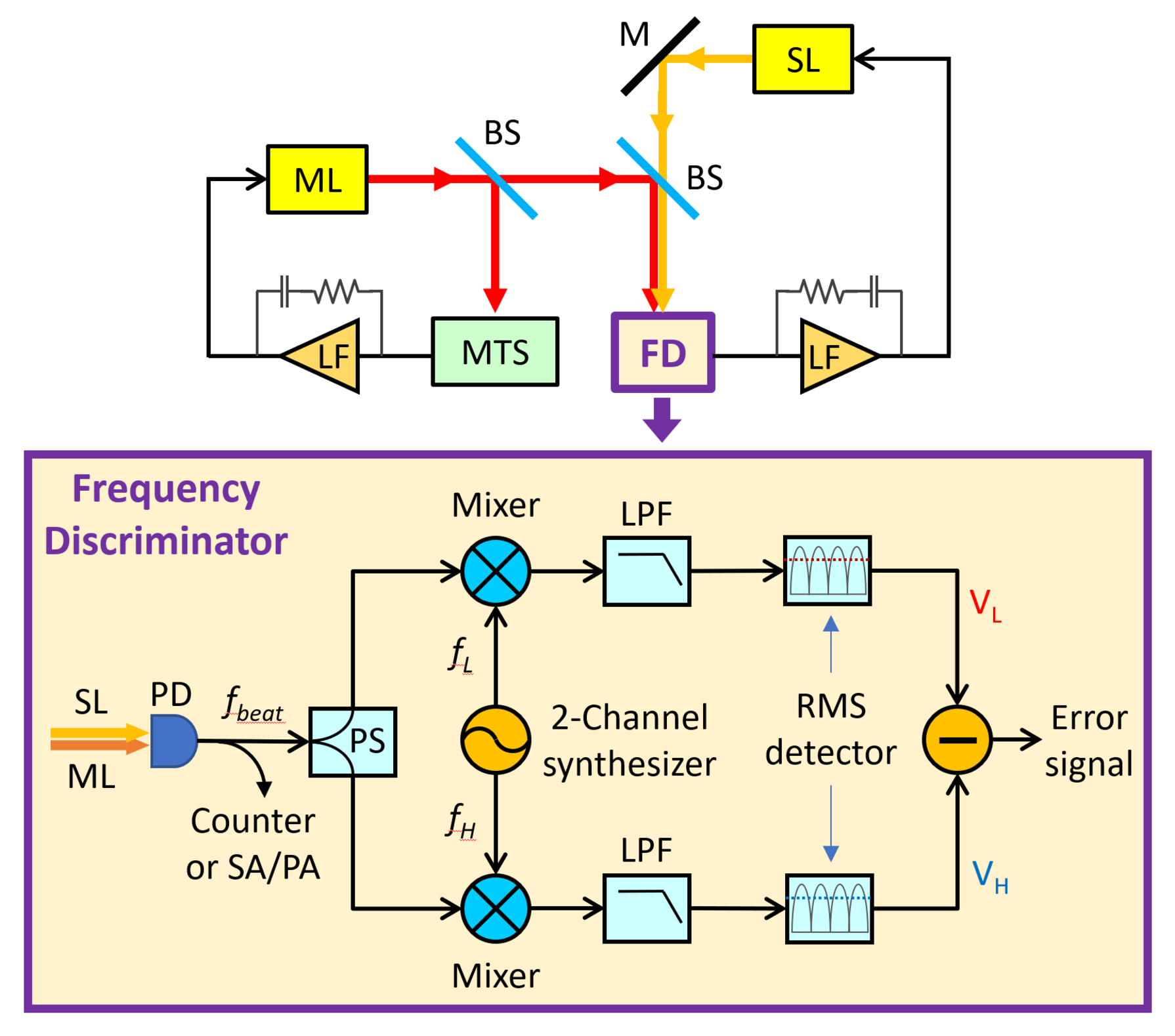}
\caption{Experimental setup and detailed schematic of the balanced filter frequency discriminator. ML: master laser; SL: slave laser; BS: beam splitter; PD: photodetector; PS: power splitter; LPF: low-pass filter; LF: loop filter; SA: spectrum analyzer; PA: phase analyzer. A 10~dB coupler is inserted between the PD and PS to tap a portion of the beat signal for diagnostic measurements.}
\label{fig:setup}
\end{figure}

The experimental setup is shown in Fig.~\ref{fig:setup}. The system uses two 852~nm external-cavity diode lasers as the master laser (ML) and slave laser (SL).\cite{Wieman1991,Park2003} The ML is frequency-stabilized to the Cs D$_2$ $6S_{1/2}$, $F=4 \rightarrow 6P_{3/2}$, $F'=5$ transition by modulation transfer spectroscopy, providing the optical reference.\cite{Park2013,Lee2023OL} The SL is offset-locked to the ML at the desired frequency difference $f_0=8.653$~GHz. This offset was chosen to generate light near the Cs D$_2$ repumping transition in our laser-cooling setup and includes the frequency shifts introduced by the acousto-optic modulators.

The optical beat note between the ML and SL is detected with a fast photodetector. A 10~dB coupler taps part of the beat signal for diagnostic measurements, while the main signal is divided into two nominally identical discriminator arms. In the two arms, the beat signal is mixed with local-oscillator frequencies $f_L=f_0-\alpha$ and $f_H=f_0+\alpha$, which are symmetrically placed around the desired offset frequency. The resulting down-converted signals are passed through low-pass filters and converted to dc voltages $V_L$ and $V_H$ using rms detectors (AD8361). The differential voltage $\epsilon=V_L-V_H$ is used as the frequency-error signal and is fed back to the SL through a loop filter.

\begin{figure}[!t]
\centering
\includegraphics[width=\columnwidth]{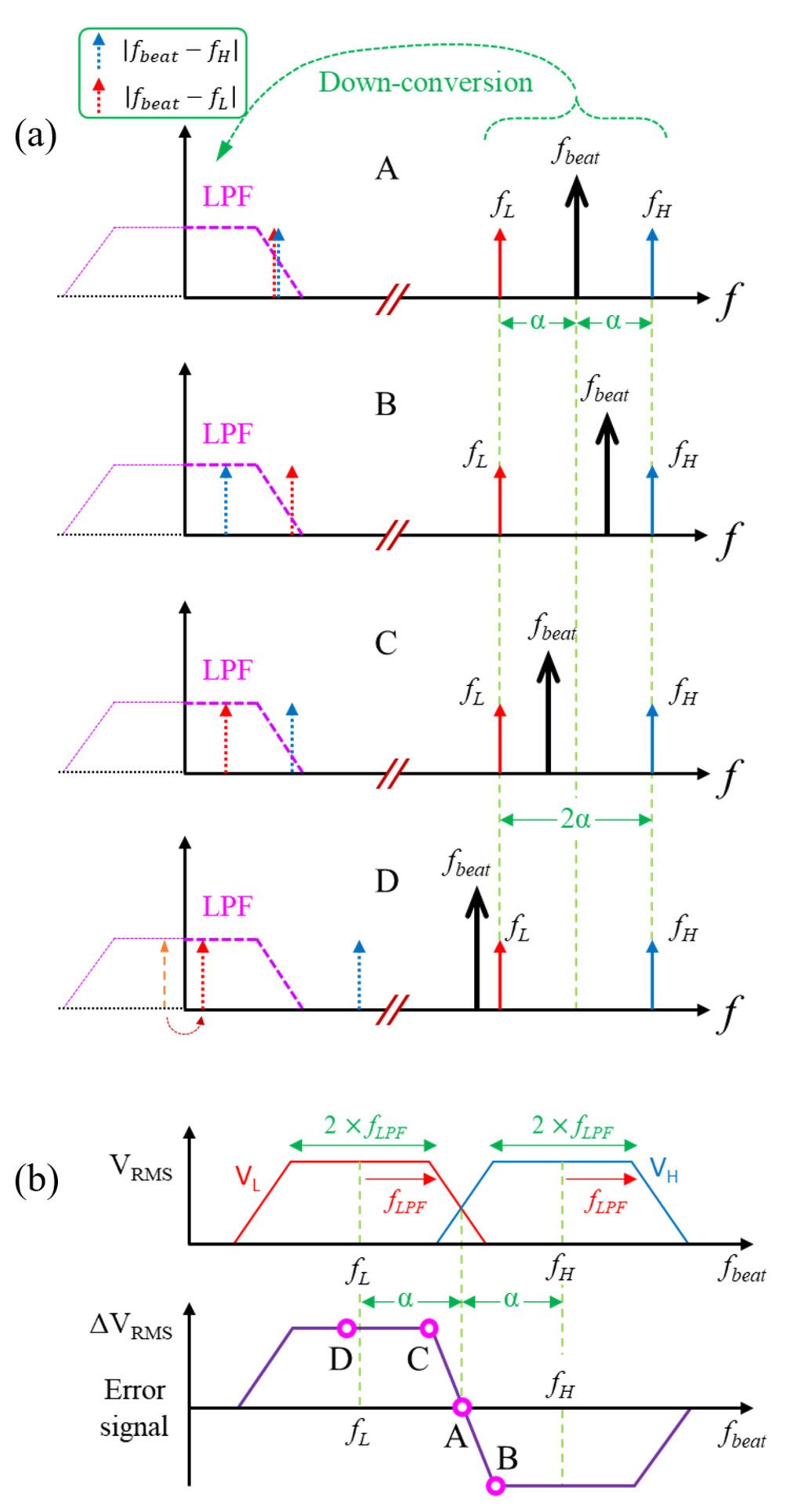}
\caption{Operating principle of the balanced filter discriminator. (a) Frequency-domain representation of the down-conversion process for cases A--D. (b) Upper: individual RMS detector voltages $V_L$ and $V_H$. Lower: resulting differential error signal showing the dispersive profile with zero-crossing points A--D.}
\label{fig:principle}
\end{figure}

\begin{figure*}[!t]
\centering
\begin{minipage}[t]{0.48\textwidth}
\centering
\includegraphics[width=\linewidth]{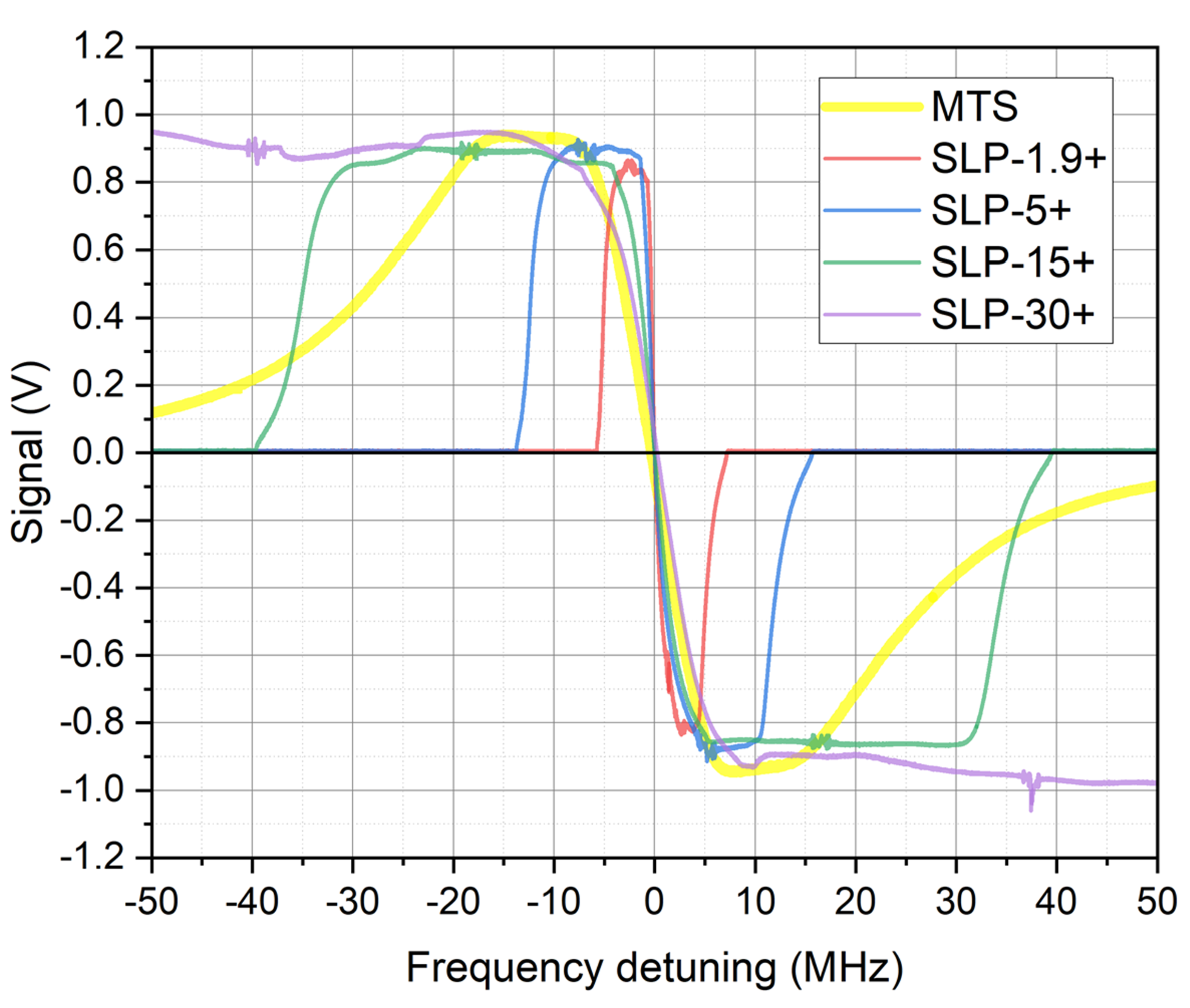}
\caption{Measured error signals for various low-pass filters (SLP-1.9+ to SLP-30+). The thick yellow line represents the MTS signal of the master laser, providing a comparative benchmark for the discrimination slopes. Narrower filters provide a steeper discrimination slope for improved sensitivity. The corresponding discrimination sensitivities are given in the text.}
\label{fig:error}
\end{minipage}\hfill
\begin{minipage}[t]{0.48\textwidth}
\centering
\includegraphics[width=\linewidth]{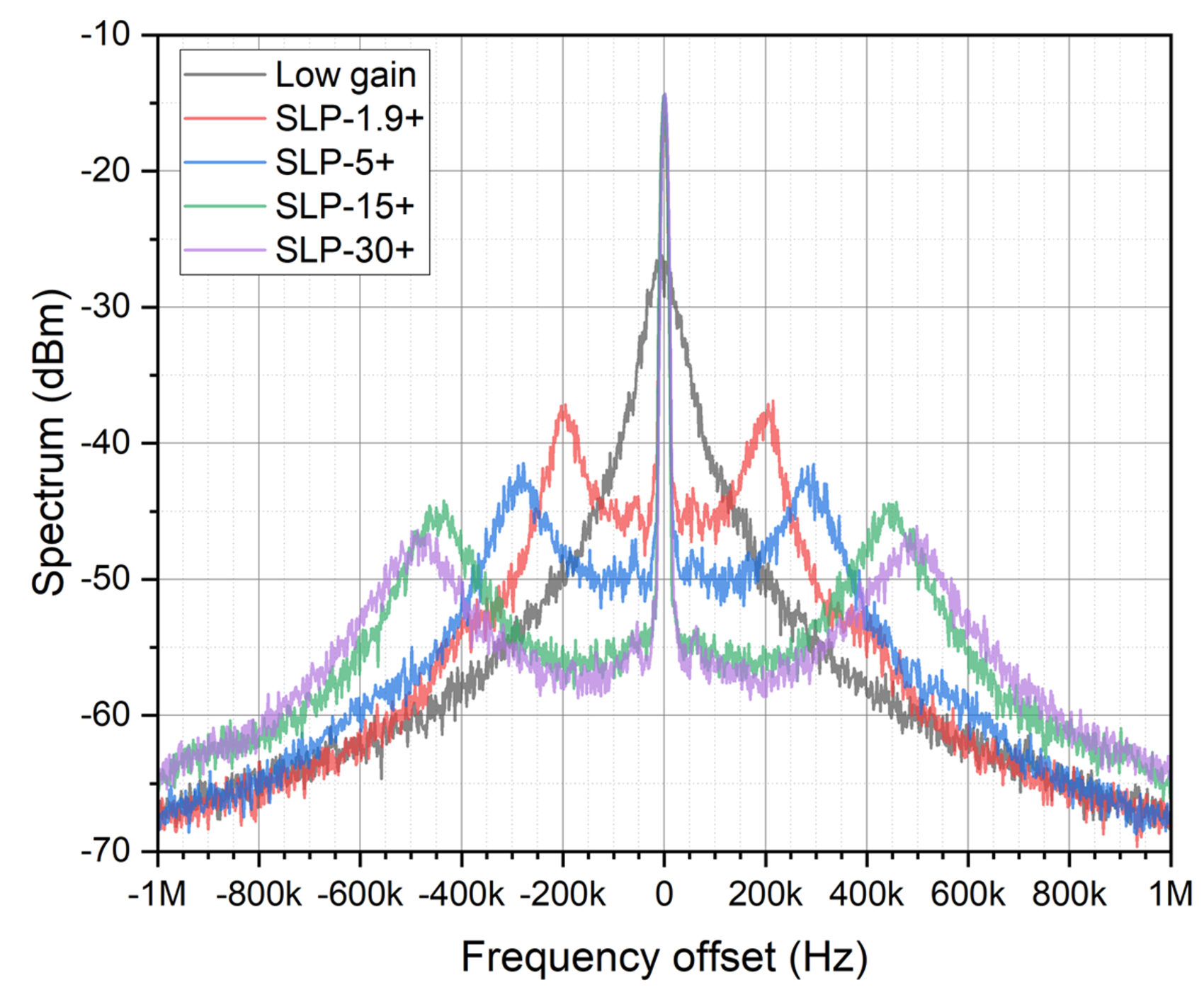}
\caption{Microwave power spectra of the 8.653~GHz beat signal under different filter configurations (RBW: 9.1~kHz). The ``Low gain'' trace shows the behavior under a marginal-gain condition.}
\label{fig:spectrum}
\end{minipage}
\end{figure*}

The operating principle of the balanced filter discriminator is illustrated in Fig.~\ref{fig:principle}. The beat frequency $f_{\mathrm{beat}}$ is compared with two local-oscillator frequencies, $f_L=f_0-\alpha$ and $f_H=f_0+\alpha$, placed symmetrically around the desired lock frequency $f_0$. After down-conversion, the two mixer-output signals are transmitted through the low-pass-filter responses of the two discriminator arms and converted into dc voltages by rms detectors. The detector outputs and differential discriminator signal can be written compactly as
\begin{align}
V_L(f_{\mathrm{beat}}) &= G_L A_L \left|H_L\left(\left|f_{\mathrm{beat}}-f_L\right|\right)\right| + V_{L,0},\nonumber\\
V_H(f_{\mathrm{beat}}) &= G_H A_H \left|H_H\left(\left|f_{\mathrm{beat}}-f_H\right|\right)\right| + V_{H,0},\nonumber\\
\epsilon(f_{\mathrm{beat}}) &= V_L(f_{\mathrm{beat}})-V_H(f_{\mathrm{beat}}).
\end{align}
Here, $H_L$ and $H_H$ are the LPF transfer functions, $A_L$ and $A_H$ are the RF amplitudes incident on the two mixer--detector chains, $G_L$ and $G_H$ are their conversion gains, and $V_{L,0}$ and $V_{H,0}$ are residual dc offsets.

In this differential configuration, common beat-amplitude fluctuations mainly produce correlated changes in the two detector outputs, whereas a frequency deviation from $f_0$ changes the two LPF transmissions in opposite directions. After simple compensation of small residual gain imbalance, the differential output provides a dispersive error signal with a zero crossing primarily defined by the midpoint of the two reference LO frequencies. The lock frequency can therefore be tuned by shifting the two LO frequencies together while preserving their separation $2\alpha$.

The four representative cases in Fig.~\ref{fig:principle}(a) summarize this behavior. In Case A, where $f_{\mathrm{beat}}=f_0$, the two IF frequencies are equal to $\alpha$, giving equal detector voltages in a balanced configuration and hence $\epsilon=0$. In Case B, $f_{\mathrm{beat}}$ is shifted above $f_0$; the IF frequency in the high-LO arm decreases toward the higher-transmission region of the LPF, while that in the low-LO arm increases toward the roll-off region, producing $V_H>V_L$ and a negative error signal. In Case C, where $f_{\mathrm{beat}}$ lies below $f_0$ but remains between $f_L$ and $f_H$, the opposite occurs and the error signal becomes positive. In Case D, for $f_{\mathrm{beat}}<f_L$, one arm approaches the low-frequency passband while the other remains far from it, leading to a saturated positive response. The corresponding differential signal in Fig.~\ref{fig:principle}(b) has a linear region around $f_0$ suitable for feedback control. Small ripples observed near the plateau regions arise from the finite low-frequency response of the mixer--detector chain and residual beat-note jitter near the zero-IF condition.

Figure~\ref{fig:error} shows the measured error signals for several Mini-Circuits SLP low-pass filters. For each LPF, the offset parameter $\alpha$ was optimized to maximize the local discrimination slope $D=d\epsilon/df$ at the zero crossing. The measured slopes were 0.12, 0.36, 0.74, and 1.48~V/MHz for the SLP-30+, SLP-15+, SLP-5+, and SLP-1.9+ filters, respectively, compared with 0.18~V/MHz for the MTS reference signal shown as a benchmark. Under the present measurement conditions, the SLP-30+ balanced-discriminator signal exhibited visibly lower noise than the MTS reference signal, despite their comparable discrimination slopes. Thus, reducing the LPF cut-off frequency increases the frequency-to-voltage conversion gain and improves the discriminator sensitivity. Small residual imbalances between the two discriminator arms were readily compensated by fine adjustment of the LO powers applied to the two mixers. Narrower filters provide higher discrimination sensitivity, whereas wider filters provide a broader capture range and lower phase delay for faster feedback. This slope--bandwidth trade-off is examined further using the locked beat spectra and Allan-deviation measurements below.

Figure~\ref{fig:spectrum} shows the microwave spectra of the 8.653~GHz beat note for different discriminator filters. Under the marginal-gain condition, where the loop gain was reduced to just above the minimum level required to maintain the lock, the beat spectrum exhibited a width of approximately 50~kHz. In contrast, the optimized locked spectra show a strong concentration of spectral power in the carrier and a reduced noise pedestal. This spectral narrowing is consistent with the enhanced effective signal-to-noise ratio of the balanced discriminator, in which common-mode amplitude fluctuations are suppressed by the differential detection.

The spectral response also reflects the slope--bandwidth trade-off discussed above. For the SLP-1.9+ configuration, servo-bump sidebands appear near $\pm 200$~kHz, indicating an effective feedback bandwidth on this order. As the LPF cut-off frequency increases, the reduced filter phase delay can support a faster feedback response. However, the feedback bandwidth does not increase proportionally for the widest LPF, indicating that the averaging capacitance of the rms detector (AD8361) can also limit the loop response. This capacitance sets the rms-to-dc averaging time and should therefore be optimized to balance detector-output ripple suppression and feedback bandwidth. Consequently, the optimum operating condition is set by both the LPF cut-off frequency and the rms-detector averaging time.

\begin{figure}[!t]
\centering
\includegraphics[width=\columnwidth]{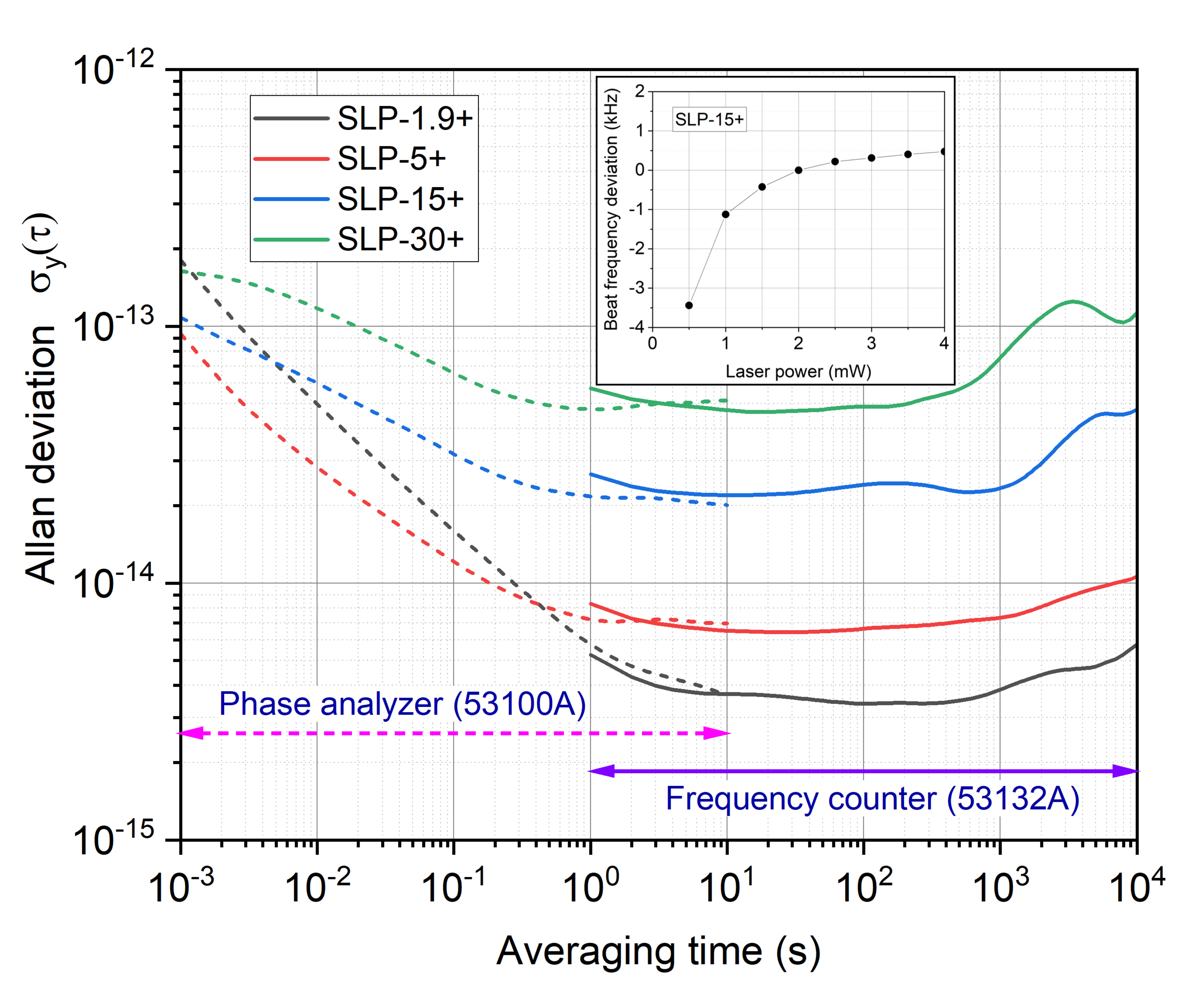}
\caption{Frequency stability evaluated by the Allan deviation $\sigma_y(\tau)$. The measured fluctuations of the 8.653~GHz beat signal are expressed as fractional frequency instability relative to the 852~nm optical carrier frequency. Data for $\tau<10$~s were obtained from phase measurements, while data for $\tau>1$~s were measured with a frequency counter; the overlapping 1--10~s range was used to confirm consistency between the two measurement methods. The SLP-1.9+ configuration achieves a stability of $4\times10^{-15}$ at 10~s. Inset: measured beat-frequency shift as a function of the optical power incident on the photodetector for the SLP-15+ configuration. The optical power was varied from 0.5 to 4~mW, and the beat-frequency excursion remained within approximately 4~kHz.}
\label{fig:allan}
\end{figure}

The frequency stability was evaluated using the Allan deviation $\sigma_y(\tau)$, as shown in Fig.~\ref{fig:allan}. The 8.653~GHz beat signal was first down-converted to the input range of the phase analyzer. Short-term fluctuations were obtained from phase measurements, avoiding the dead time associated with conventional frequency counting, whereas long-term data were recorded with a high-resolution frequency counter. The overlapping 1--10~s region confirmed consistency between the two measurement methods. No linear drift was removed before calculating the Allan deviations. The measured beat-frequency fluctuations were expressed as fractional frequency instability relative to the 852~nm optical carrier.

The Allan-deviation curves show that the narrower filters provide better long-term stability because of their larger discrimination slopes, while the wider filters show a tendency toward slightly improved stability at short averaging times, consistent with their faster feedback response. The SLP-1.9+ configuration, which provides the largest measured discrimination slope, achieved $\sigma_y(1~\mathrm{s})=5\times10^{-15}$ and reached $4\times10^{-15}$ at 10~s when referred to the 852~nm optical carrier. The instability remained below approximately $6\times10^{-15}$ up to $10^4$~s. This value is lower than the $3\times10^{-14}$ instability at 10~s reported for one of the most stable frequency-only offset locks based on hybrid electronic filters.\cite{Li2022}

The inset of Fig.~\ref{fig:allan} shows the dependence of the locked beat frequency on the optical power incident on the photodetector. For the representative SLP-15+ configuration, the beat-frequency excursion remained within approximately 4~kHz as the optical power was varied from 0.5 to 4~mW. Although a finite residual power-dependent frequency shift remains, the small observed excursion shows that the optimized differential configuration substantially suppresses amplitude-to-frequency conversion. In this implementation, the residual balance was adjusted simply by fine tuning the LO powers applied to the two mixers.

We have demonstrated a high-stability laser offset-frequency locking system based on a balanced filter discriminator. The discriminator uses two down-conversion arms driven by local-oscillator frequencies placed symmetrically around the desired offset frequency. After low-pass filtering and RMS detection, differential subtraction of the two detector outputs produces a dispersive frequency-error signal with a zero crossing primarily defined by the reference local-oscillator frequencies. This configuration provides a practical frequency-only offset-locking method while reducing sensitivity to common beat-power fluctuations.

The system was implemented for an 8.653~GHz offset between two 852~nm external-cavity diode lasers, corresponding to the frequency difference used for cesium repumping in our laser-cooling setup. By comparing several low-pass filters, we observed the trade-off between discrimination sensitivity and feedback bandwidth. The SLP-1.9+ configuration reached a fractional frequency instability of $4\times10^{-15}$ at 10~s and remained below approximately $6\times10^{-15}$ up to $10^4$~s, when referred to the 852~nm optical carrier. The measured power-dependent beat-frequency shift was limited to approximately 4~kHz over a photodetector optical-power range of 0.5--4~mW. This result shows that the optimized differential configuration effectively suppresses amplitude-to-frequency conversion.

This balanced-discriminator approach is useful for atomic-physics laser systems requiring stable and tunable frequency offsets, including laser cooling, repumping, and optical pumping. Because the lock point is set by the midpoint of the two LO frequencies, the offset frequency can be tuned accurately by shifting $f_L$ and $f_H$ together while keeping $\alpha$ fixed, without relying on a calibrated dc offset voltage. The method can also be adapted to other offset ranges through suitable down-conversion or frequency-division stages, although such stages may affect the achievable stability. In frequency-comb--referenced systems, selected comb modes may serve as master references for stabilizing auxiliary lasers with large optical-frequency separations, including potentially THz-scale offsets, when combined with suitable down-conversion or frequency-division stages. These results show that the balanced filter discriminator provides a practical and stable frequency-only offset-locking method for atomic-physics and precision-frequency-metrology applications.

\begin{acknowledgments}
This work was supported by the Korea Research Institute for Defense Technology Planning and Advancement (KRIT) grant funded by the Defense Acquisition Program Administration (DAPA) (KRIT-CT-23--031) and Measurement Technology for Grand National Strategic Industries funded by the Korea Research Institute of Standards and Science (KRISS 2026-GP2026--0012).
\end{acknowledgments}

\end{document}